\documentclass{iaus}
\usepackage{graphicx}

\title[S264.~~Solar cycles : the past evolution influence] 
{Solar cycles : the past evolution influence}

\author[Alexis Klutsch \& Rubens Freire Ferrero]   
{Alexis KLUTSCH$^{1,2}$
 \and Rubens FREIRE FERRERO$^2$}

\affiliation{$^1$Universidad Complutense de Madrid, Departamento de Astrof\'isica, 
Facultad C.C. F\'isicas, 28040 Madrid, Spain, email: {\tt klutsch@astrax.fis.ucm.es} \\[\affilskip]
$^2$Observatoire Astronomique, Universit\'e de Strasbourg \& CNRS, UMR 7550, 11 rue de l'Universit\'e, 67000 Strasbourg, France, email: {\tt rubens.freire@astro.unistra.fr}}

\pubyear{2009}
\volume{264}  
\pagerange{xxx--xxx}
\jname{Solar and Stellar Variability Impact on Earth and Planets}
\editors{A.~H.~Andrei, A.~Kosovichev, J.-P.~Rozelot, eds.}
\begin{document}

\maketitle

\begin{abstract}
The so-called \textsf{solar cycle} is generally characterized by the quasi-periodic oscillatory evolution of the photospheric spots number. This quasi-periodic pattern has always been an intriguing question. Several physical models were proposed to explain this evolution and many mathematical data analysis were employed to determine the principal frequencies noticeable in the measured data. Both approaches try to predict the future evolution of the solar activity and to understand the physical phenomena producing these cycles. Here we present the analysis of the sunspots number evolution using the time-delay approach. Our results show than the solar cycle can also be characterized by this behavior implying the influence of the past evolution over the present one, suggesting an histeresis mechanism, linked probably with magnetic activity.

\end{abstract}

\firstsection 

\section{Introduction}

Solar activity can be seen through the evolution of sunspots number in quasi-periodic oscillatory series with periods going from 8 to 15 years and with a mean period of 11~years. Due to the change of magnetic field polarity in solar hemispheres alternatively each cycle, the period is rather 22 years. The quasi-periodic evolution of 
many activity phenomena 
is still a unsolved key problem in solar physics (along with, e.g., heating of the solar chromosphere and corona, and solar flares). An important issue of this understanding is due to the influence of solar activity over the terrestrial climate (\cite[Archibald 2006]{arch06}). 

\medskip
Many physical models were proposed to understand the basic mechanisms (e.g., \cite[Benevolenskaya 1998]{bene98}) producing solar/stellar activity and its quasi-cyclic evolution. 
From sunspots time series, mathematical approaches have highlighted several other hidden periods other that the well-apparent 11-years period. These works contributed to provide some observed parameters to constraint theoretical models, and to predict the future evolution of the sunspots number (e.g., \cite[Clilverd \etal\ 2003]{clil03}; \cite[Sello 2003]{sello03}). 

\medskip
In this work we follow another mathematical approach, investigating the influence of the past evolution of the sunspots number over the present one. A natural and simple way to carry out this analysis is to assume that the susnpots number have a temporal delay behaviour. Thus, the relationship between present and past values becomes non-linear and could prove that the past amplitude influence the present one. As for the abovementioned methods, this one also allows some predictions about future solar cycles by using intermediate parameters to characterize the general past evolution. 

 \vspace*{-0.3 cm}
\section{The solar cycle as a temporal delay phenomena}

We consider a temporal delay behavior (Eq.~\ref{Eq1}) to rely the present evolution of a phenomena with some particular events in its past evolution. The variation of $N$ in time is not only related with its current value $N(t)$ but also with its past values $N(t-T)$, where $T$ is our temporal delay. The simplest way is to assume only one past temporal influence and some proportionality between the variables (e.g., \cite[Murray 1993]{mur93}), as follows: 

\begin{equation}
	\frac{\textit{d}N(t)}{\textit{d}t} = a ~ N(t) \left\{ 1 - b ~ N(t-T) \right\} ~~\to~~ \frac{\Delta N_{i}}{\Delta t} = a ~ N_{i} \left\{ 1 - b ~ N_{i-J} \right\}
       \label{Eq1}
\end{equation}
 
\begin{figure}
    \begin{minipage}{0.35\textwidth}   
 \caption{\scriptsize  \textit{Left panel:} Correlation factor obtained by chi-square minimization versus the temporal delay. \textit{Right panel:} Sunspots number ratio versus sunspots number with a 8-years delay. The blue line shows a linear fit.}
   \label{Fig:RatioFactor}
    \end{minipage}
    \hfill
    \begin{minipage}{0.65\textwidth}
    \centerline{\includegraphics[width= 3.2in]{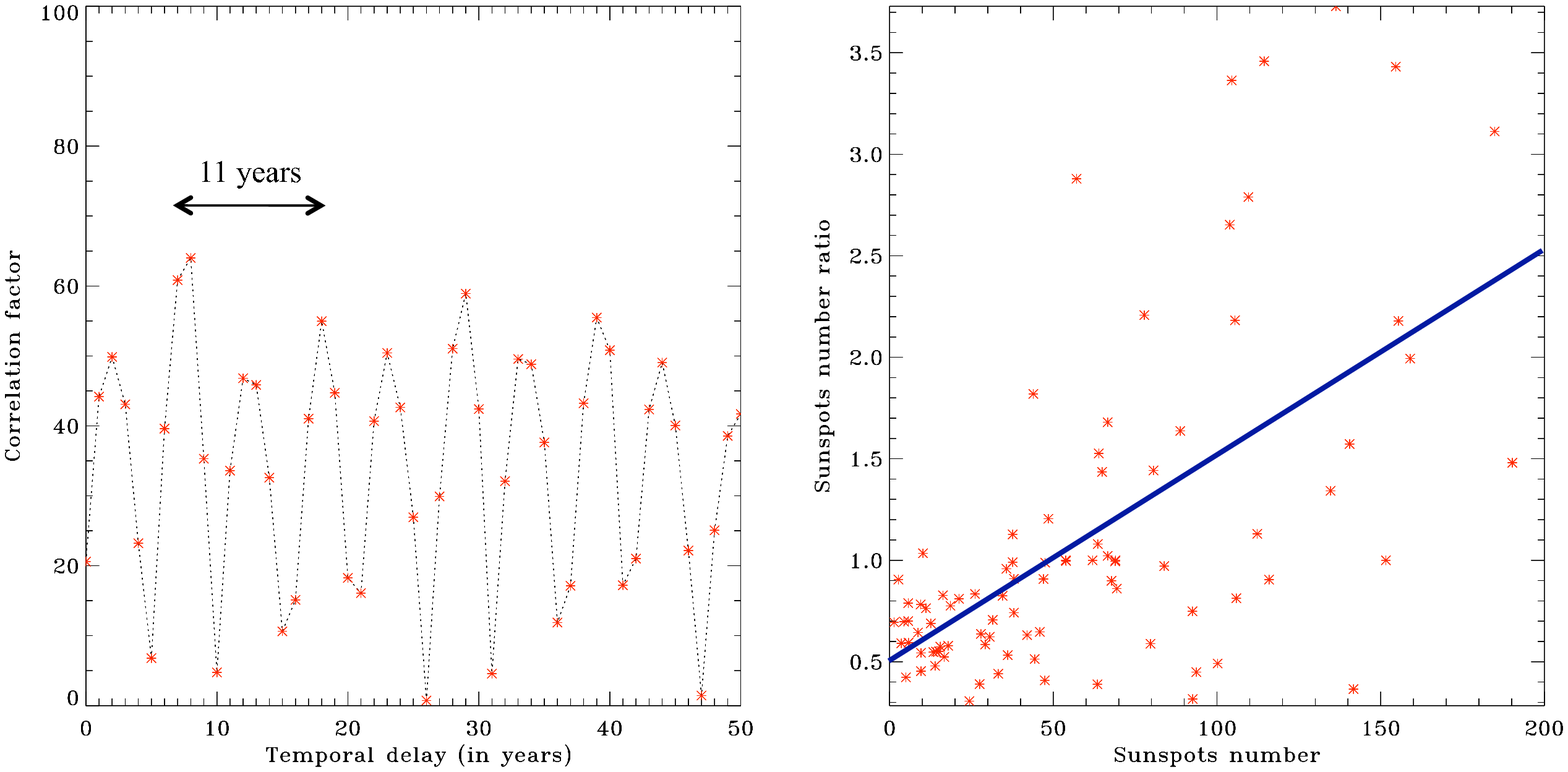}}
    \end{minipage}
    \end{figure}

Because we use yearly values of the relative sunspots number, given by SDIC (http:// 
www.sidc.be/sunspot-data/), $\Delta t$ is equal to $1$. The present and past values of $N$ are already known and we assumed some values for the parameter $T$ ($T = 0,1,2, \dots$; Fig.~\ref{Fig:RatioFactor}, left panel). So we can determine constants $a$ and $b$ (Eq.~\ref{Eq2}) from the correlation between the sunspots number ratio and the past sunspots number (Fig.~\ref{Fig:RatioFactor} right panel).  

\begin{equation}
	N_{i} + \Delta N_{i}           = a ~ N_{i} \left\{ 1 - b ~ N_{i-J} \right\}  + N_{i} ~~\to~~ \frac{N_{i+1}}{N_{i}}                          =  \left\{ a + 1 \right\} - a b ~ N_{i-J}
       \label{Eq2}
\end{equation}

 \vspace*{-0.5 cm}
\section{Results  and conclusions}

\begin{figure}[h]
\begin{center}
 \includegraphics[width=1.6in]{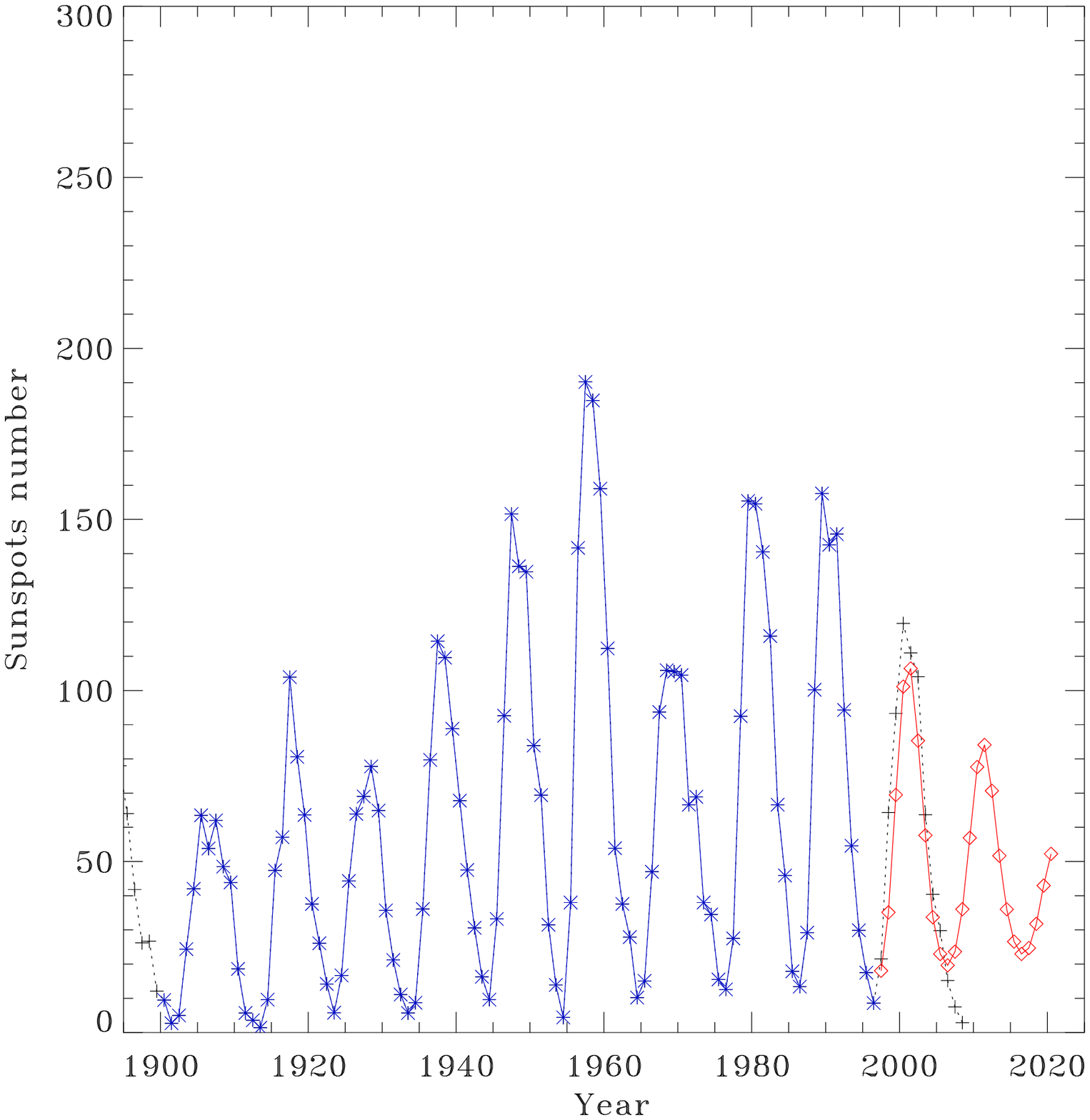} 
 \includegraphics[width=1.55in]{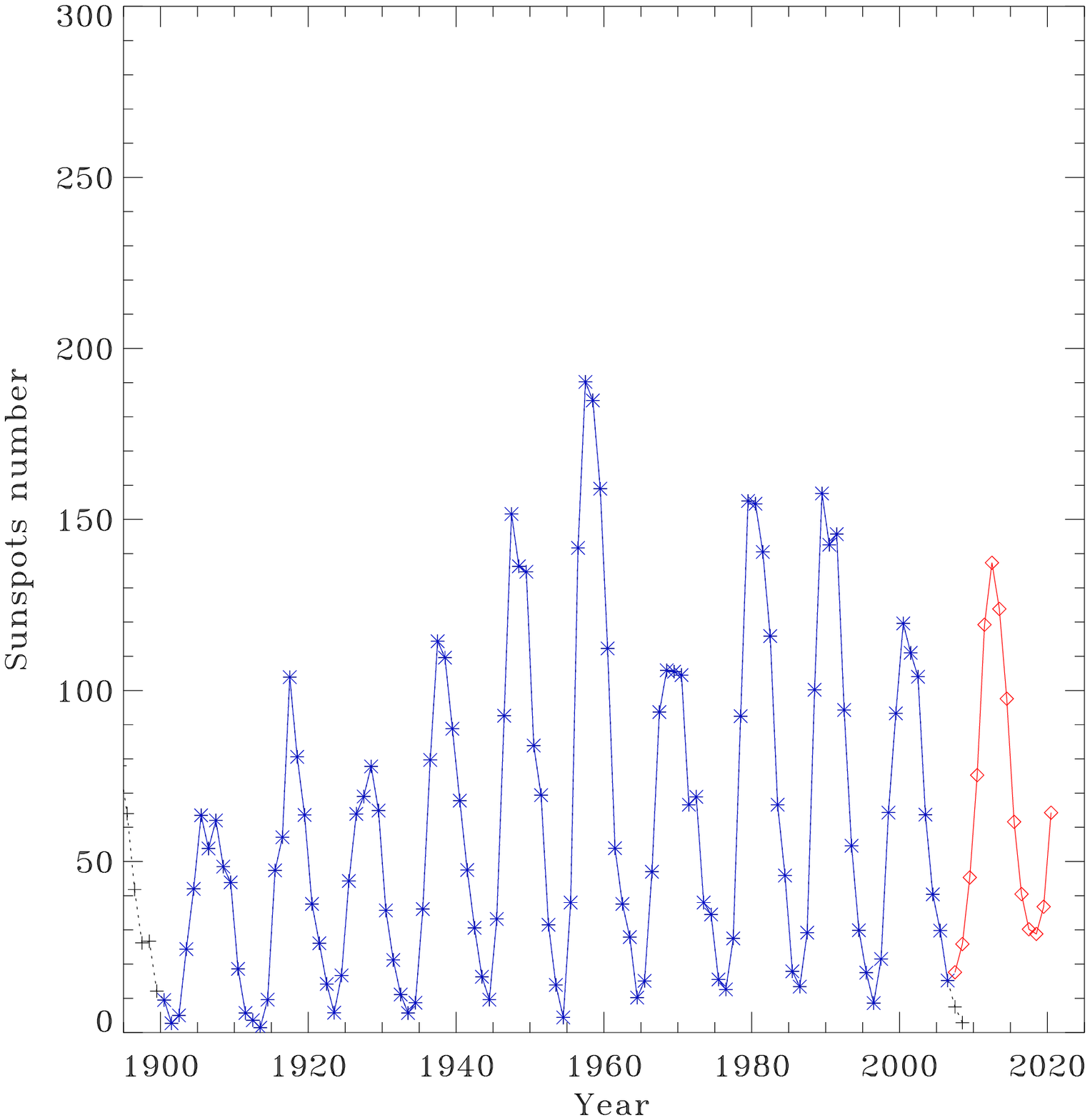} 
 \includegraphics[width=1.55in]{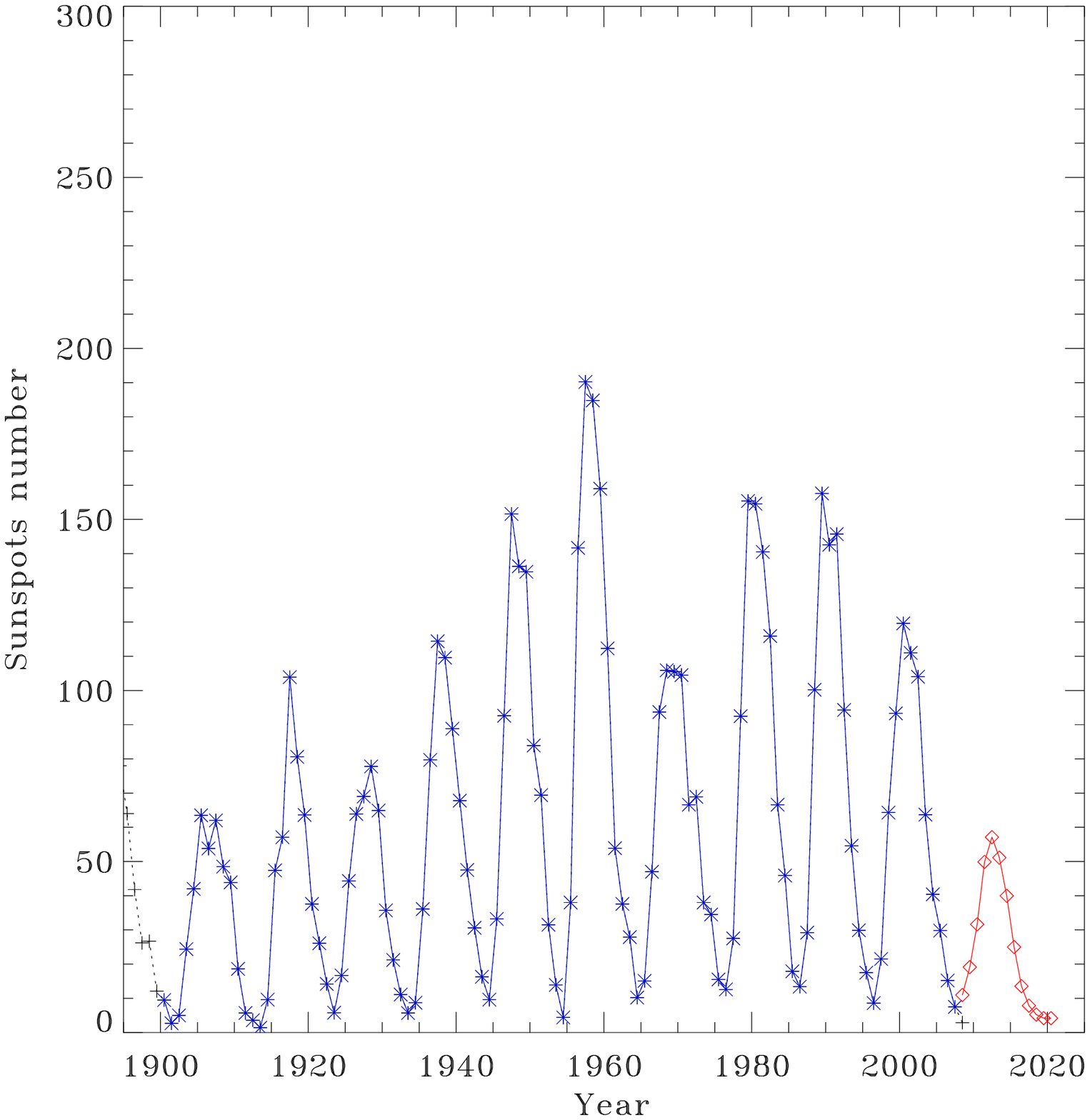} 
 \caption{\scriptsize The observed (\textit{plus symbols and dotted line}) and selected (\textit{asterisks and blue line}) yearly values of the relative sunspots number. Our predictions using the temporal delay method are also marked (\textit{diamonds and red line}). \textit{Left panel:} Prediction for the solar cycle 23 and 24. \textit{Middle and Right panels:} Prediction of the solar cycle~24 using the sunspots number observed until 2006.5 and 2007.5 as input.}
   \label{Fig:EvolSpots}
\end{center}
\end{figure}

\begin{figure}[h]
 \vspace*{-0.6 cm}
    \begin{minipage}{0.65\textwidth}   
    \caption{\scriptsize Maximum ($R_{max}$) versus minimum ($R_{min}$) sunspots number of a given solar cycle (\textit{dots}). The cycle number is also marked. Blue, green and red diamonds show the locus of our predicted maxima using our modelling of the solar cycle~$23$ as well as the spots number observed until $2006.5$ and $2007.5$, respectively (Table~\ref{Tab:Prediction}). The predicted range of \cite[Brajsa \etal\ (2009)]{Brajsa09} is plotted as well (\textit{vertical line}). The blue and red lines show linear fits using data of both axis and the linear least-square fit like one given by \cite[Brajsa \etal\ (2009, Fig.~6)]{Brajsa09}, respectively.}
    \end{minipage}
    \hfill
    \begin{minipage}{0.35\textwidth}

 \vspace*{0.35 cm}
    \centerline{ \includegraphics[width= 1.6in]{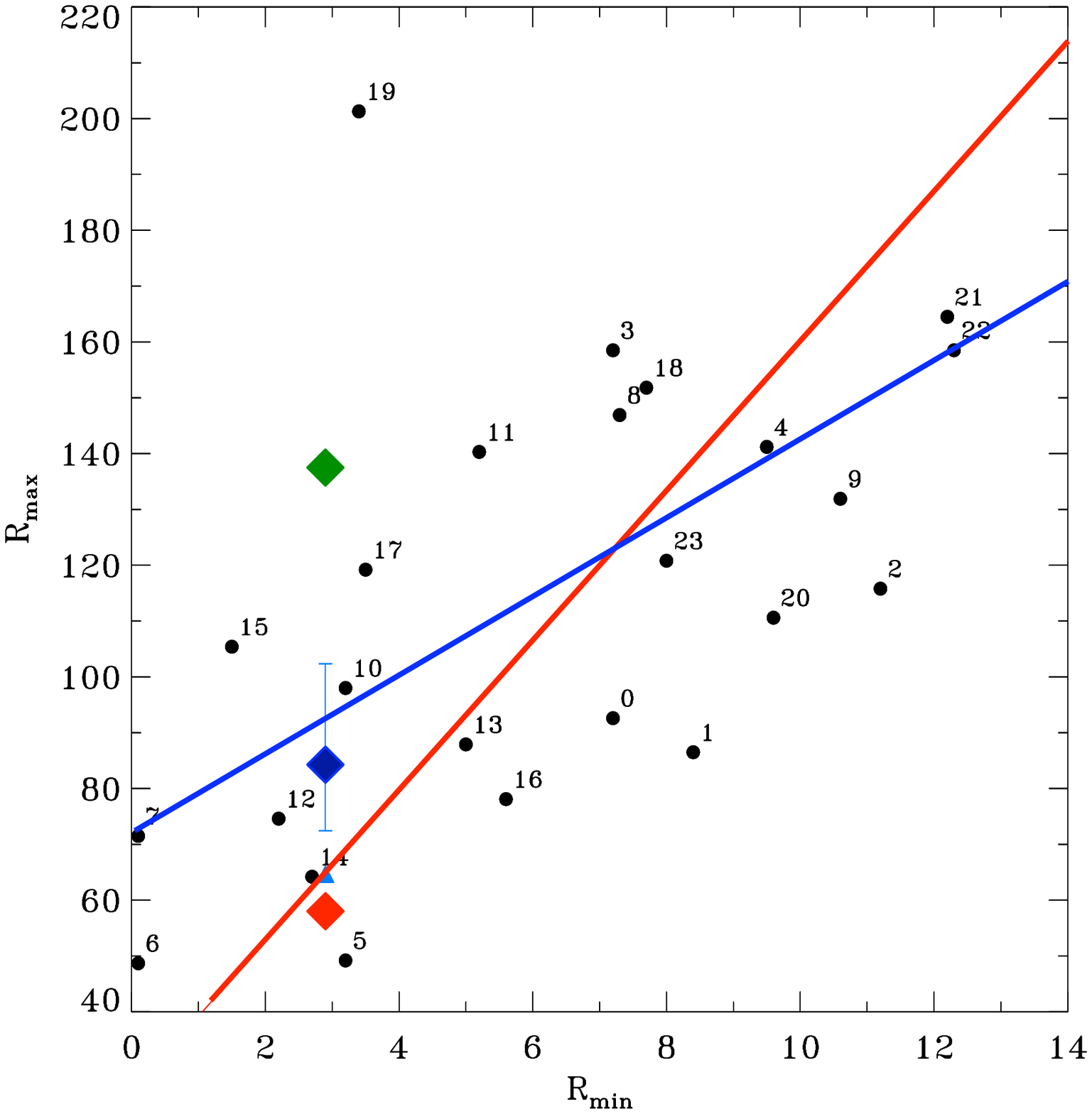}}
    \label{Fig:RminRmax}
    \end{minipage}
    \end{figure}

The $T$ values which minimize the correlation between the sunspots number ratio and the past spots number, are 7 or 8 years (Fig.~\ref{Fig:RatioFactor}). To test the accuracy of our method, we applied it on the solar cycle~23 (Table~\ref{Tab:Cycle23} and Fig.~\ref{Fig:EvolSpots}, left panel). Our predictions are consistent with the observations, except for the unusually long period of low activity at the end of this cycle. We also characterized the next solar cycle~$24$ (Table~\ref{Tab:Prediction} and Fig.~\ref{Fig:EvolSpots}) using our modelling of the solar cycle~$23$ as well as the spots number observed until $2006.5$ and $2007.5$. The next maximum sunspots number would take place between $2011$ and $2012$ and should be close to 60 (Fig.~\ref{Fig:EvolSpots}, right panel) as for the solar cycle~$14$. Using the solar minimum occurred in $2008.5$ ($R_{min} = 2.9$ from SDIC), the predicted maximum is consistent with the value from the linear least-square fit on the both axis (red line on Fig.~\ref{Fig:RminRmax}) and is a few smaller than the range given by \cite[Brajsa \etal\ (2009)]{Brajsa09}. Moreover the next solar minimum should not occur before 2019 or even 2020.

\begin{table}[h]
  \begin{center}
  \caption{ {\scriptsize Observed and predicted values for the cycle 23. Results showed in Fig.~\ref{Fig:EvolSpots} (left panel).}}
  \label{Tab:Cycle23}
 {\scriptsize
  \begin{tabular}{cccccccc}
\noalign{\medskip}
  \hline 
  \multicolumn{2}{l}{Observed parameters} & \multicolumn{2}{c}{Epoch of solar} & \multicolumn{2}{c}{Maximum sunspot} & \multicolumn{2}{c}{Epoch of the end} \\
  \multicolumn{2}{l}{of solar minimum}       & \multicolumn{2}{c}{maximum}        & \multicolumn{2}{c}{number}                 & \multicolumn{2}{c}{of cycle}              \\
\noalign{\medskip}
  Epoch                & Number                    & Observed        & Predicted            & Observed        & Predicted                & Observed        & Predicted               \\
  \hline 
  $1996.4$            & $8.0$                        & $2000.3$        & $2001.5$            & $120.8$           & $106.4$                   & $2009.?$        & $2006.5$                \\
  \hline 
\noalign{\medskip}
  \end{tabular}
  }
 \end{center}
\end{table}

\begin{table}[h]
  \begin{center}
  \caption{\scriptsize Predicted values for the solar cycle 24 using various data set in input.}
  \label{Tab:Prediction}
 {\scriptsize
  \begin{tabular}{lcccccc}
\noalign{\medskip}
  \hline 
  Panel of                        & \multicolumn{2}{c}{Input parameters of} & \multicolumn{2}{c}{Predicted parameters} & \multicolumn{2}{c}{Predicted parameters of} \\
  Fig.~\ref{Fig:EvolSpots} & \multicolumn{2}{c}{solar minimum}        & \multicolumn{2}{c}{of solar maximum}      & \multicolumn{2}{c}{next solar minimum}  \\
\noalign{\medskip}
                & Epoch                & Number                      & Epoch                & Number                    & Epoch                & Number  \\
  \hline 
  Left        & $2006.5$            & $19.6$                        & $2011.5$            & $~~84.1$                 & $2016.5$             & $23.0$  \\
  Middle    & $2006.5$            & $15.2$                        & $2012.5$            & $137.3$                   & $2018.5$             & $28.8$  \\
  Right      & $2007.7$            & $~~7.5$                      & $2012.5$            & $~~57.1$                 & $2020.5$             & $~~4.2$  \\
  \hline 
\noalign{\medskip}
  \end{tabular}
  }
 \end{center}
\end{table}

These preliminary results are encouraging because we find a similar delay as that observed between the geomagnetic activity and solar cycle peaks \cite[(Hathaway \& Wilson 2006)]{hat06} linked probably with magnetic activity by some kind of histeresis mechanism. At present, we can successfully reproduce some previous solar cycles (e.g., epoch and solar maximum). We plan to improve our predictive method including the influence of delays for which there is a good correlation. This more detailed description could allow to solve the overestimation of the next solar minimum and to reproduce the asymmetries observed during the previous cycles. Our final aim is to obtain a reliable prediction of the whole solar activity by identifying the fundamental lower and high activity precursors of its present and future evolution. All these improvements are still needed to be able to predict future low solar activity periods (e.g., Maunder/Dalton Minimum). 



\begin{thebibliography}{}




 {\scriptsize
\bibitem[Archibald (2006)]{arch06}
{Archibald, D.} 2006, Energy \& Environment, 17, 29

\bibitem[Benevolenskaya (1998)]{bene98}
{Benevolenskaya, E.~E.} 1998, \textit{ApJ}, 509, L49

\bibitem[Braj{\v s}a \etal\ (2009)]{Brajsa09}
{Braj{\v s}a, R., W{\"o}hl, H., Hanslmeier, A., et al.} 2009, \textit{A\&A}, 496, 855

\bibitem[Clilverd \etal\ (2003)]{clil03}
{Clilverd, M.~A, Clarke, E., Rishbeth, H., et al.} 2003, \textit{Astronomy and Geophysics}, 44, 5.20

\bibitem[Hathaway \& Wilson (2006)]{hat06}
{Hathaway, D.~H. \& Wilson, R.~M.} 2006, Geophys. Res. Lett., 33, L18101

\bibitem[Murray (1993)]{mur93}
{Murray, J.~D.} 1993,  \textit{Mathematical Biology, 2nd corr. ed.} (Springer)

\bibitem[Sello (2003)]{sello03}
{Sello, S.} 2003, \textit{A\&A}, 410, 691
}
\end{thebibliography}
\end{document}